\documentclass[fleqn,usenatbib]{mnras}

\usepackage{newtxtext,newtxmath}

\usepackage[T1]{fontenc}

\usepackage{multirow}
\usepackage{amsmath}
\usepackage{xcolor}

\definecolor{Red}{rgb}{0.743,0,0}
\definecolor{Blue}{rgb}{0.25,.41,.88}
\definecolor{Green}{rgb}{0,0.5,0}

\def\deg{\ifmmode {^\circ}\else {$^\circ$}\fi}
\def\degree{\ifmmode {^\circ}\else {$^\circ$}\fi}
\def\mum{\ifmmode {\rm \,\mu {\rm m}}\else $\rm \,\mu {\rm m}$\fi}
\def\arcsec{\ifmmode ^{\prime \prime}\else $^{\prime \prime}$\fi}

\def\inch{\ifmmode ^{\prime \prime}\else $^{\prime \prime}$\fi}
\def\msunyr{\ifmmode {M_{\odot}~{\rm yr^{-1}}}\else $M_{\odot}~{\rm yr^{-1}}$\fi}
\def\msun{\ifmmode {M_{\odot}}\else $M_{\odot}$\fi}
\def\rsun{\ifmmode {R_{\odot}}\else $R_{\odot}$\fi}
\def\lsun{\ifmmode {L_{\odot}}\else $L_{\odot}$\fi}
\def\mstar{\ifmmode {M_{\star}}\else $M_{\star}$\fi}
\def\rstar{\ifmmode {R_{\star}}\else $R_{\star}$\fi}
\def\tstar{\ifmmode {T_{\star}}\else $T_{\star}$\fi}
\def\lstar{\ifmmode {L_{\star}}\else $L_{\star}$\fi}
\def\md{\ifmmode {M_d}\else $M_d$\fi}
\def\ld{\ifmmode {L_d}\else $L_d$\fi}
\def\ad{\ifmmode A_d\else $A_d$\fi}
\def\ldlstar{\ifmmode L_d / L_\star\else $L_d / L_{\star}$\fi}
\def\rearth{\ifmmode {\rm R_{\oplus}}\else $\rm R_{\oplus}$\fi}
\def\mearth{\ifmmode {\rm M_{\oplus}}\else $\rm M_{\oplus}$\fi}
\def\qdstar{\ifmmode Q_D^\star\else $Q_D^\star$\fi}
\def\kms{\ifmmode {\rm km~s^{-1}}\else $\rm km~s^{-1}$\fi}
\def\ms{\ifmmode {\rm m~s^{-1}}\else $\rm m~s^{-1}$\fi}
\def\mesc{\ifmmode m_{esc}\else $m_{esc}$\fi}
\def\rmin{\ifmmode r_{min}\else $r_{min}$\fi}
\def\rmax{\ifmmode r_{max}\else $r_{max}$\fi}
\def\mmin{\ifmmode m_{min}\else $m_{min}$\fi}
\def\mmax{\ifmmode m_{max}\else $m_{max}$\fi}
\def\rmind{\ifmmode r_{min,d}\else $r_{min,d}$\fi}
\def\rmaxd{\ifmmode r_{max,d}\else $r_{max,d}$\fi}
\def\mmaxd{\ifmmode m_{max,d}\else $m_{max,d}$\fi}
\def\vrad{\ifmmode v_{rad}\else $v_{rad}$\fi}
\def\qz{\ifmmode q_{0}\else $q_{0}$\fi}
\def\qi{\ifmmode q_{i}\else $q_{i}$\fi}
\def\ql{\ifmmode q_{l}\else $q_{l}$\fi}
\def\qs{\ifmmode q_{s}\else $q_{s}$\fi}
\def\rbrk{\ifmmode r_{brk}\else $r_{brk}$\fi}
\def\rdamp{\ifmmode r_{damp}\else $r_{damp}$\fi}
\def\rin{\ifmmode r_{in}\else $r_{in}$\fi}
\def\rout{\ifmmode r_{out}\else $r_{out}$\fi}
\def\tin{\ifmmode t_{in}\else $t_{in}$\fi}
\def\tout{\ifmmode t_{out}\else $t_{out}$\fi}
\def\ain{\ifmmode a_{in}\else $a_{in}$\fi}
\def\aout{\ifmmode a_{out}\else $a_{out}$\fi}
\def\r0{\ifmmode R_{0}\else $R_{0}$\fi}
\def\m0{\ifmmode m_{0}\else $m_{0}$\fi}
\def\M0{\ifmmode M_{0}\else $M_{0}$\fi}
\def\xm{\ifmmode x_{m}\else $x_{m}$\fi}
\def\sigz{\ifmmode \Sigma_0\else $\Sigma_0$\fi}
\def\gyr{\ifmmode {\rm g~yr^{-1}}\else ${\rm g~yr^{-1}}$\fi}
\def\cms{\ifmmode {\rm cm~s^{-1}}\else ${\rm cm~s^{-1}}$\fi}
\def\gcms{\ifmmode {\rm g~cm^{-2}}\else $\rm g~cm^{-2}$\fi}
\def\gcmss{\ifmmode {\rm g~cm^{-2}~s^{-1}}\else $\rm g~cm^{-2}~s^{-1}$\fi}
\def\gcmc{\ifmmode {\rm g~cm^{-3}}\else $\rm g~cm^{-3}$\fi}
\def\dcm2{\ifmmode {\rm dyn~cm^{-2}}\else $\rm dyn~cm^{-2}$\fi}
\def\ecsk{\ifmmode {\rm erg~cm^{-1}~s^{-1}~K^{-1}}\else $\rm erg~cm^{-1}~s^{-1}~K^{-1}$\fi}
\def\cm2{\ifmmode {\rm cm^{-2}}\else $\rm cm^{-2}$\fi}
\def\atilin{\ifmmode {\tilde{a}_{in}}\else $\tilde{a}_{in}$\fi}
\def\atilout{\ifmmode {\tilde{a}_{out}}\else $\tilde{a}_{out}$\fi}
\def\atil{\ifmmode {\tilde{a}}\else $\tilde{a}$\fi}
\def\ttil{\ifmmode {\tilde{t}}\else $\tilde{t}$\fi}
\def\sqrttt{\ifmmode {\tilde{t}^{1/2}}\else $\tilde{t}^{1/2}$\fi}

\def\h2o{H$_2$O}
\def\sio2{SiO$_2$}
\def\ch4{CH$_4$}
\def\h2{H$_2$}
\def \ms{m\,s$^{-1}$\,}
\def \kms{km\,s$^{-1}$}
\def \msun{M$_{\odot}$}
\def \rsun{R$_{\odot}$}
\def \lsun{L$_{\odot}$}

\def \mearth{M$_{\oplus}~$}
\def \me{M$_{\oplus}$}

\DeclareRobustCommand{\VAN}[3]{#2}
\let\VANthebibliography\thebibliography
\def\thebibliography{\DeclareRobustCommand{\VAN}[3]{##3}\VANthebibliography}

\usepackage{graphicx}	
\usepackage{amsmath}	

\title[Loss of an atmosphere by planetesimal impacts
]{About the loss of a primordial atmosphere of super-Earths by planetesimal impacts}

\author[Lozovsky et al.]{
Michael Lozovsky,$^{1}$\thanks{E-mail: michloz@mail.com}
Dina Prialnik,$^{1}$
and Morris Podolak$^{1}$
\\
$^{1}$Department of Geosciences,\\ Tel-Aviv University, Tel-Aviv,\\ Israel
}

\date{Accepted XXX. Received YYY; in original form ZZZ}

\pubyear{2022}

\begin{document}
\label{firstpage}
\pagerange{\pageref{firstpage}--\pageref{lastpage}}
\maketitle

\begin{abstract}
We consider planets composed of water ice and rock, located  far from a central star. In an earlier study, computing the growth of planets by continuous accretion, we found that a large fraction of the ice evaporates upon accretion, creating a water vapor atmosphere. Here we consider accretion as a discrete series of planetesimal impacts (of order $10^8$), at the same time-dependent accretion rate, and investigate the fate of the vapor, as a result of its interaction with the accreting planetesimals. We find that a large fraction of the vapor escapes. The remaining fraction may form an outer layer of ice after the termination of accretion and cooling of the surface. The escaped water mass may significantly alter the ice-to-rock ratio of the planet.  We investigate the effect of different choices of parameters such as the ice-to-rock ratio, the planetesimal size distribution, and the impact velocities. We find that the planetesimal size distribution has a negligible effect and explain why. By contrast, the ice-to-rock ratio and impact velocities affect the fraction of retained water masses considerably. 
\end{abstract}




The study of planetary formation has been undergoing a transformation in recent years. As the classic core accretion  paradigm predicts, rocky planets form by accretion of solids from a circumstellar protoplanetary disk \citep{Pollack1996}. Their composition and internal structure depend on the disk's chemical composition and solid surface density, which are determined by the location of the planet and the disk's physical and chemical properties. 

It is common to assume that ice-rich Super-Earths were formed exterior to the water ice line \citep{Lodders2003,Ros2013,Marcus2010}. However, the details of the structure and distribution of the different chemical compounds is a subject of debate. While some models predict homogeneous interiors \citep{Vazan2022}, others argue that a layered structure is formed, with a rocky core below a water ice layer \citep[e.g.][]{Lozovsky2022,Dorn2021,Lozovsky2018}. 

This outermost ice layer of a rocky protoplanet might be subject to sublimation, and the consequent formation of a primordial water vapor atmosphere. However, this initial water vapor atmosphere might be partly subject to mass loss by irradiation \citep[e.g.][]{Howe2020} and more important might be striped as a consequence of impacts \citep{Schlichting2015}. As the planet forms, it is subject to the impacts of solid planetesimals of various sizes, which can be as large as few thousand kilometers in diameter \citep[e.g.][and references within]{Johansen2014,Guilera2014}. Those planetesimal impacts can cause an additional atmospheric loss via various mechanisms: First, the impact itself may cause a shock wave in the atmosphere, that will eject part of its mass; secondly, the falling mass heats up the atmosphere by deposition of potential energy. This heating leads to additional evaporation of the gas.

In this study, we follow the formation of an ice-rich Super-Earth, with emphasis on the fate of its primordial water atmosphere. Instead of using a continuous function to describe the accretion rate, as in \citet{Lozovsky2022} (hereafter: Paper I), we simulate the accretion by a series of discrete planetesimal impacts, where the planetesimal masses are chosen from a given size distribution. The method of calculation is described in Section~\ref{sec:method} and the results are described and discussed in Section~\ref{sec:results}.

\section{Method of calculation}
\label{sec:method}

\subsection{Motivation}
\label{subsec:Moti}

\begin{figure}
 	\centering
	\includegraphics[width=0.45\textwidth, trim={4.0cm 9.5cm 2.9cm 9.5cm},clip]{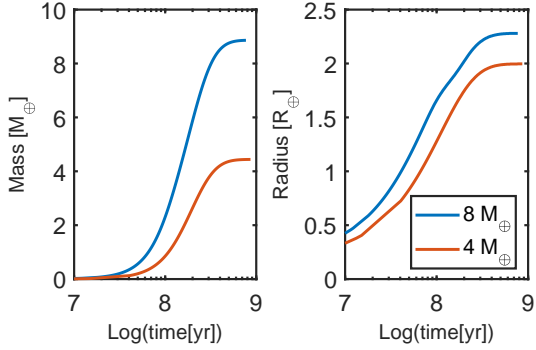}
	\caption{Formation of a planet of 8 \me~ (blue line) and 4 \me~ (red line). Both models  assume a 30\% ice mass fraction for a planet formed at an orbital distance of 40 au. Models from Paper I. }
	\label{fig:Formtion}
\end{figure}

Paper I followed the formation of an 8\me~ and a 4\me ~ ice-rich planets, surrounding a 1\msun ~ star at an orbital distance of 40 au. The structure of the forming planet depends strongly on the assumed initial conditions. In the simulation, the accretion rate $\dot M$ was computed according to \cite{Lissauer1987}:

\begin{equation}\label{eq:saf}
	\dot{M}(t)=\frac{\sqrt{3}\pi}{2} R(t)^2\Sigma_s(t) \Omega \left[ 1+2\Theta\right],
\end{equation}
where $R$ is planetary radius, $\Sigma_s$ is solid surface density of the disk, $\Omega$ is the Keplerian frequency and $\Theta=v_{esc}^2/2v^2$ is the Safronov parameter \citep[see][]{safronov1972}, $v_{esc}$ is the escape velocity from the accreting body, and $v$ is the random velocity of a planetesimal. The resulting planet's growth shown in Fig.~\ref{fig:Formtion} is described in detail in Paper I. 

We found in Paper I that during the accretion process, ice near the surface heated by accretion energy, evaporated. The loss of vapor into space lowered the resulting ice to rock ratio in the planet.
In that study we assumed that any ice that evaporated was entirely lost from the system. However, the thermal energy alone of the evaporated water is insufficient to overcome the gravitational energy well of the protoplanet. Therefore a primordial water atmosphere is expected to form. The question we now address is, how much of the evaporated water would be really lost and how much, retained.

The accretion of planetesimals not only supplies additional water to the planet and heats the surface sufficiently for water to sublimate, but it also supplies additional kinetic energy to the atmospheric gas that might allow it to escape.  One particular mechanism of mass loss due to a planetesimal impact was studied in detail by \cite{Schlichting2015}.  Adopting the results of that work and implementing them to a long series of planetesimal impacts, we calculate the outcome of the full growth of a planet by planetesimal accretion, based on the time-dependent accretion rate that we obtained in Paper I. Our focus is on the escape of the vapor that evaporates upon accretion.

\subsection{Atmosphere removal by planetesimals}
\label{ss:removal}

Assume the protoplanet has an exponential atmospheric density profile such that the density, $\rho$ as a function of height above the surface, $z$, is given by
\begin{equation}
\rho=\rho_0 \cdot exp[-z/h],
\end{equation}
where $h$ is the atmospheric scale height. If the planetesimal is more massive than $m\gtrsim \sqrt{2}\rho_0(\pi hR)^{3/2}$, the planetesimal will eject all of the atmosphere above the tangent plane at the point of impact \citep{Schlichting2015}.  Bodies smaller than this but larger than $m=4\pi\rho_0h^3$, will eject only a fraction of the atmospheric mass above the tangent plane.  Using self-similar solutions for the hydrodynamic equations of shock propagation, \cite{Schlichting2015} were able to derive the following equation for the fraction of the planet's atmosphere, $f$, that is removed by a single impactor:

\begin{equation}
  f =
    \begin{cases}
      0.4x +1.4x^2 -0.8 x^3 \text{ (for isothermal atmosphere)},\\
      0.4x +1.4x^2 -1.2 x^3 \text{ (for adiabatic atmosphere)}.
    \end{cases}\label{eq:f}       
\end{equation}

 Here $x=v_{imp}m/v_{esc}M$, where $m$ and $v_{imp}$ are impactor mass and relative velocity, respectively, $v_{esc}$ is the protoplanet's escape velocity and $M$ is the protoplanet's mass at the time of impact. As $v_{imp}\approx v_{esc}$ (explained later), $x \approx m/M$.
 
\begin{figure}
 	\centering
	\includegraphics[width=0.95\columnwidth, trim={0.3cm 0.3cm 0.1cm 0.1cm},clip]{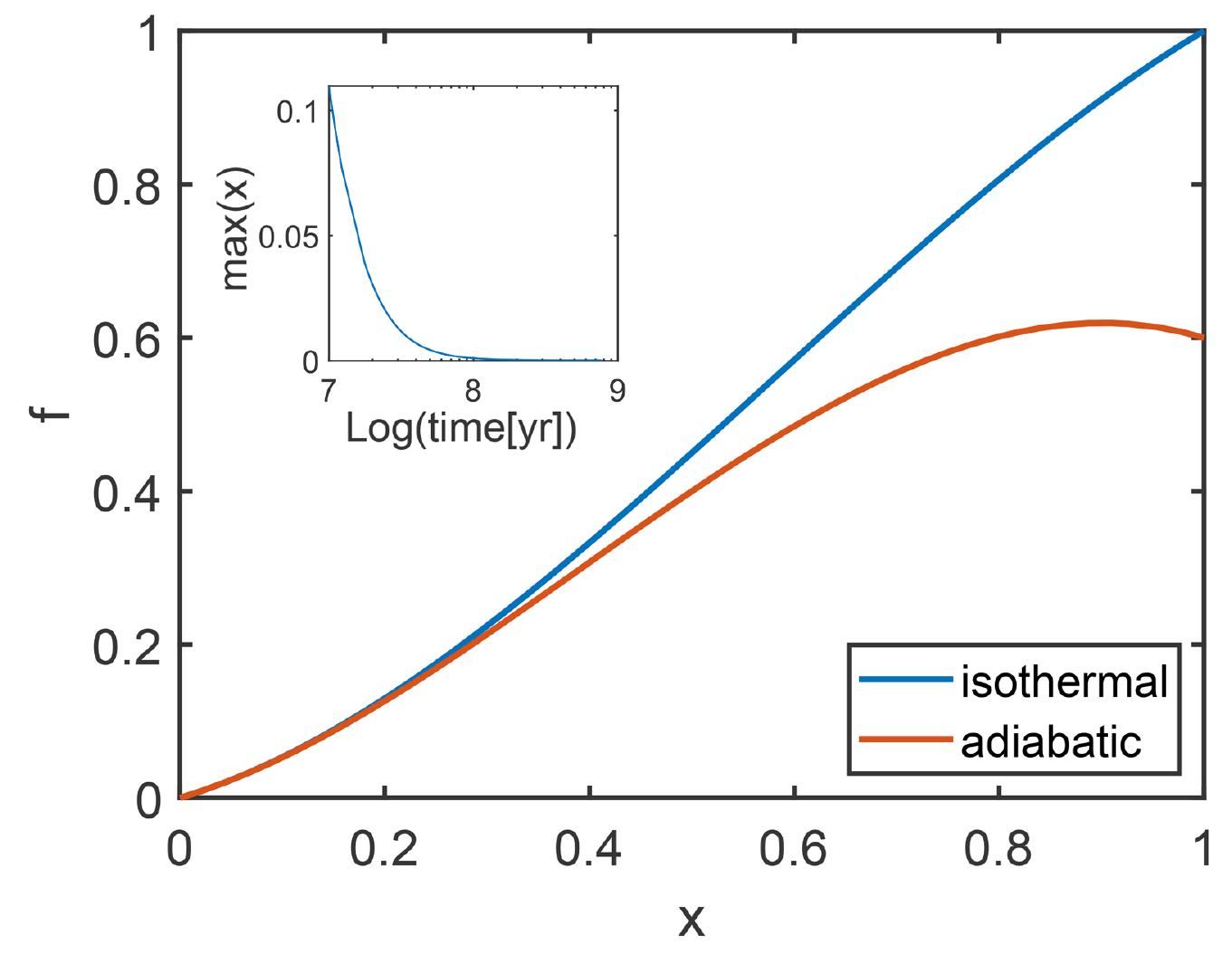}
	\caption{The fraction $f$ of atmospheric mass removed by an impact, as a function of the impact parameter $x=v_{imp}m/v_{esc}M$, form Eq.~\ref{eq:f}. The maximal $x$ versus time is shown in a small inset figure.} 
	\label{fig:Xloss}
\end{figure}

The difference between the two cases is shown in Fig.~\ref{fig:Xloss}. We note that if the planetesimals are small relative to the protoplanet, there is almost no difference in $f$ between the the isothermal and adiabatic regimes. In our study, the maximal possible size of a planetesimal is very small compared to the protoplanet after $10^7 $ yrs. Therefore the two versions of $f(x)$ in Eq.~(\ref{eq:f})  yield almost identical results.

\subsection{Planetesimal size distribution}
\label{distribution}

We assume that accretion takes place by a series of planetesimal collisions, where---for a given case---all planetesimals have the same density $\rho_p$ and the same ice to rock ratio, so that a planetesimal is defined by its radius $a$. The radii,  divided into a total of $I$ size bins, are chosen to have radii defined by a geometric series in the range $[a_{\rm min},a_{\rm max}]$. The geometric factor $q$ is determined by the number of size bins according to
\begin{equation}
\log{q}=\frac{\log{(a_{\rm max}/a_{\rm min})}}{I-1}.
\end{equation}
 
We adopt a discrete power law size distribution for the radii $a_i$ ($1 \le i \le I$), with a power index $\gamma$.  The total mass of the initial planetesimal distribution is equal to the final planetary mass $M_p$; the number $N_i$ of planetesimals of radius $a_i$ is thus given by
\begin{equation}
N_i=\frac{3M_p}{4\pi \rho}\frac{a_i^{-\gamma}}{\sum_i a_i^{3-\gamma}}.
\end{equation}

We define $S$ as the total number of planetesimals, $S=\sum_1^I N_i$ arranged as a string, and choose a random number $j$ in the range $[1,S]$. Its location in the string determines the corresponding size bin $i$ of the planetesimal that will impact the planet. We assume that the entire mass $m_i=4\pi a_i^3\rho/3$ is added to the planet. Consequently, the number of planetesimals in bin $i$ decreases by 1, and a new string is defined of range $[1,S-1]$, from which another planetesimal is randomly chosen,  and so on. 


\subsection{How does it work?}
\label{ss:work}

We start with the results of evolutionary calculations based on a continuous mass accretion rate (Paper I). The input data is a series of vectors of properties corresponding to discrete times during evolution. For each time $t_k$ ($1\le k \le K$), we have $M_k$ - the planetary mass accumulated up to that time, $R_k$ - the planetary radius, and $A_k$ - the mass of vapor ejected by the planet up to that time. 
With this data, we define $\Delta M_k = M_k-M_{k-1}$ and $\Delta A_k=A_k-A_{k-1}$ for each $1< k \le K$, where $\Delta M_1=M_1$ and $\Delta A_1=A_1$. The final mass of the planet is $M_p=M_K$ and the total mass of ejected vapor is $A_v=A_K$.

Consider the time interval between $k$ and $k+1$ in the evolution data. The accreted mass during this interval is $\Delta m_{k+1}=M_{k+1}-M_k$. We choose planetesimals randomly from the distribution until their total mass $\Delta m_{\rm imp}=\sum_j m_j$ exceeds $\Delta m_k$.  We adjust the input data vector, so that at the new point $k+1$, the corresponding mass matches the total mass of planetesimals that have been accreted. 
 Since each impact removes a fraction $f_j$ of the atmosphere (see Section~\ref{ss:removal}), the remaining fraction is $1-f_j$, so that consecutive impacts reduce the atmosphere by a factor
  \begin{equation}
  f_k=\Pi_j (1-f_j).
  \end{equation}
Therefore, if the mass of the atmosphere at time $k$ is $B_k$, at time $k+1$ it will be
 \begin{equation}
 B_{k+1}=f_k(B_k+\Delta A_{k+1}).
\end{equation}  

Eventually, a bin---say $i_1$---will be emptied of planetesimals. We remove this bin, and shift up all bins above $i_1$. In the end we are left with one bin and it, too, will be emptied. This will mark the end of planetesimal accretion.

The results of each evolution run depend on several free parameters. Three parameters determine the distribution of planetesimal sizes:
 the size range of planetesimals, that is, the choice of each $a_{\rm min}$ and $a_{\rm max}$; the power law $\gamma$, which determines the relative number of large impactors; and
the number of size bins $I$, which determines the resolution (accuracy) of the procedure.
Two additional parameters determine the planetesimal properties and the impact intensity:
the impact velocity, more precisely, the ratio $v_{\rm imp}/v_{\rm esc}$, which is determined by Safronov parameter; and the ice-to-rock ratio of the planetesimals. 
We have tested the effect of each of these parameters on the evolution path and on the final atmospheric mass.



\section{Results and conclusions}
\label{sec:results}

We define a benchmark model of $\sim$ 8\me, which is formed by planetesimals with a composition of 30\% ice and 70\% rock by mass. The default range of planetesimal sizes is chosen to be 20-1000 km. The time-dependent accretion rate adopted for the cases considered in this study is obtained for a normalized Safronov parameter $\bar{\theta}=400$ (see Paper I), corresponding to a Safronov parameter of $\Theta=16000$.
 The Safronov parameter determines the random planetesimal velocity, and hence the impact velocity, as $\Theta= v^2_{esc}/2v^2$ and $v_{imp}=\sqrt{v_{\infty}^2+v_{esc}^2}$. The high value of $\Theta$ means that $v_{esc}$ is large compared to $v$ and leads to impact velocities very nearly equal to the escape velocity $v_{esc}$. This planetesimal impact velocity is used for our benchmark model.
 
 \begin{figure*}
 	\centering
	\includegraphics[width=0.78\textwidth, trim={0.0cm 1.6cm 0.5cm 2.1cm},clip]{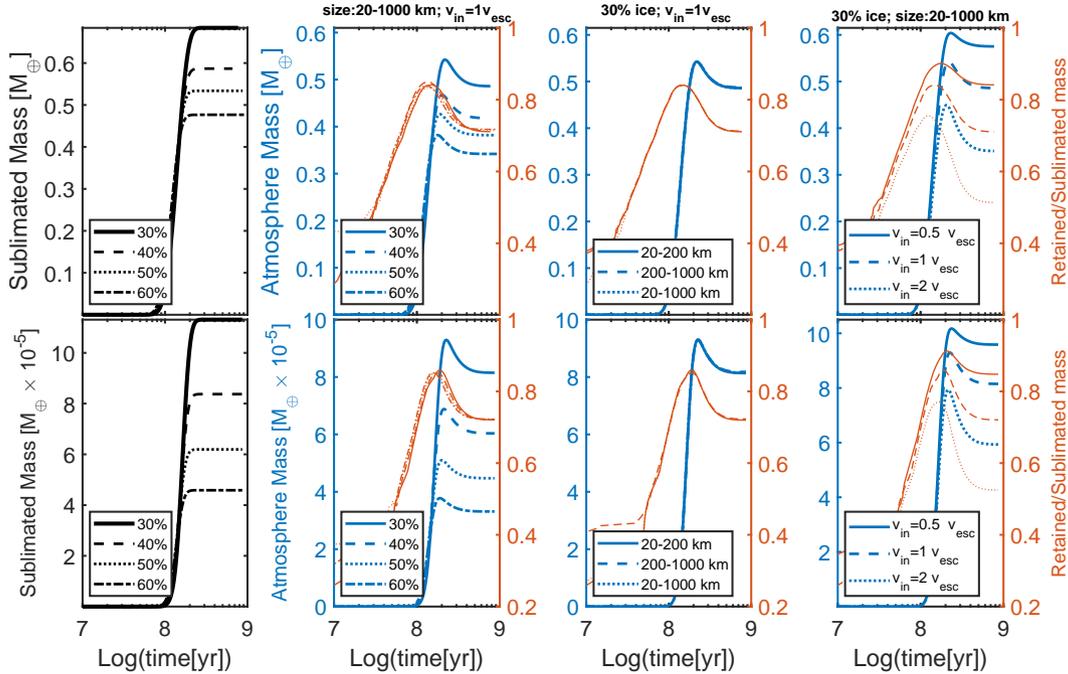}
	\caption{Formation of the atmosphere for $M_p \sim 8 M_{\oplus}$ (upper panels), and $M_p \sim4 M_{\oplus}$ (lower panels). The leftmost panels show the total sublimated mass, taken from Paper I. The other panels show the retained atmospheric mass after the impacts (blue) and the ratio of retained to sublimated mass (red). The second couple of panels show results for various ice mass fractions, the next two panels, the effect of various planetesimal ranges and the rightmost panels, the effect of various impact velocities.}
	\label{fig:Vars}
\end{figure*}

We tested various ice mass fractions, planetesimal size ranges and impact velocities, as shown in Fig.~\ref{fig:Vars}. The four upper and lower panels show the formation of a $\sim 8$\me ~and a $\sim 4$\me~ planet, respectively. The leftmost panels show the sublimated mass (black curves) and we note that it depends strongly on the assumed ice-to-rock ratio, as explained in Paper I. The crucial factor, however, is the planetary mass: reducing the total mass by half leads to a decrease by four orders of magnitude in the atmospheric mass.

The second upper and lower panels show that the atmospheric mass follows the trend of the evaporated mass: a higher ice mass fraction results in a lower-mass atmosphere. One should note that for different ice mass fractions, the retained/sublimated mass ratios (red lines) diverge slightly in the first stages, but they converge by the end of the process. In the third upper and lower panels, testing the effect of different palenetesimal size distributions, we find only minor differences and only in the first stages of accretion. Thus, the ratio of retained/sublimated mass converges to the same value, $\sim 0.72$, regardless of either ice content or size distribution of the planetesimals.

This convergence might look surprising at first glance, but it can actually be predicted analytically. Each impact reduces the atmospheric mass by $[1-f(x)]$ and by the time the main growth of the planet starts, $x$ is already small (see Fig.~\ref{fig:Xloss}), and according to Eq.~(\ref{eq:f}), $f\approx0.4x$, where $x=m/M$ (see Section~\ref{ss:removal}). For simplicity, we may assume that all planetesimals are of the same size, hence after $n$ impacts, $M=n m$ and thus $x_n=1/n$. The ratio of atmospheric mass to sublimated masss is therefore given by

\begin{multline}
\Pi_n (1-f(x_n) ) \approx \Pi_n\left(1-\frac{0.4}{n}\right)\approx
\left(1-\frac{0.4}{n}\right)^n \\
\rightarrow {\rm e}^{-0.4}=0.67
\label{eq:analytic}
\end{multline}
which is quite close to the results we get from the detailed simulations.

We also tested systematically the effect of the size distribution parameters, by varying each parameter in turn (the number of size bins $I$, the power law $\gamma$, the minimal size, keeping the benchmark maximal size, and the maximal size, keeping the minimal benchmark size, as well as the seed of the random number generator) assuming extreme values, and comparing the results with the benchmark model. The differences in retained atmospheric mass are less than $2\times10^{-3}$. As predicted by \cite{Schlichting2015}, the cumulative effect of many low-mass impactors is matched by the effect of a few massive ones; this is why the size distribution is far less important than the total mass of the impactors (i.e., the planetary mass).

Only the different impact velocities can change the retained atmospheric mass (as shown in rightmost two panels), with higher impact velocity leading to lower retained mass. Although as a test, we allowed the impact velocity to vary, we should note that the Safronov parameter determines the impact velocity, and therefore most of the planetesimals are expected to have impact velocities of $\sim v_{esc}$. We note that for $v_{imp}/v_{esc}>1$, the value of $x$ increases by the same factor, and hence the convergence limit decreases.

The peak in atmospheric mass that appears in all cases, corresponds to the decline of sublimation, which occurs after the maximum of the accretion rate (when accretion heating weakens); thereafter planetesimal impacts continue to remove atmospheric mass, until the end of accretion, after which the atmospheric mass remains constant.

 \begin{figure}
 	\centering
	\includegraphics[width=0.41\textwidth, trim={3.5cm 8.5cm 3.10cm 7.9cm},clip]{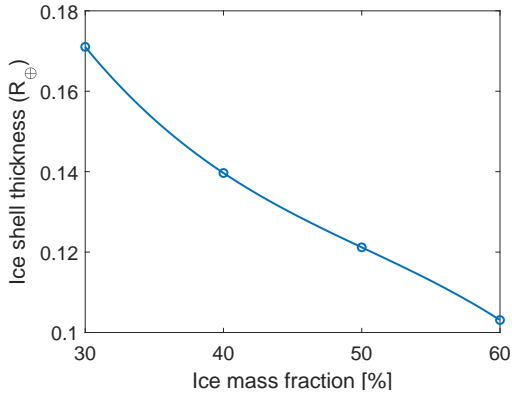}
	\caption{Ice shell thickness for models with different initial ice mass fractions, assuming the retained ice to condense on the surface.}
	\label{fig:IceShell}
\end{figure}

In conclusion, our results show that the falling planetesimals are able to strip a large fraction of the primordial water vapor atmosphere, while keeping part of it intact. As the planet is located in a cold environment, the water vapor is expected to cool down when accretion ends, and form a pure ice layer. Assuming that all the retained water mass condenses down and solidifies and taking the ice density $\rho_{ice}$ at low temperatures as  constant and equal to 0.934~gr/cm$^3$, the thickness of the ice layer can be calculated from the volume $V=M_{atm}/\rho_{ice}=\frac{4}{3}\pi (R^3-R_{p}^3)$, where $R_{p}$ is the planetary radius without the ice shell (as calculated in Paper I), $R$ is the radius of the planet with the pure ice layer. The thickness of this layer $z_{ice}=R-R_{p}$ is shown in Fig.~\ref{fig:IceShell}.

To summarize, while the planetesimal impacts deliver volatile material that forms a primordial atmosphere, the same planetesimals themselves also strip it off by impacts. However, a large fraction of the ice is retained, and cools to form a new icy layer.
We suggest that the accretion scenario presented here should be considered when modeling the formation of Super-Earths that include a significant fraction of ice. 

\section*{Acknowledgements}

We thank Christian Reinhardt for valuable discussions and suggestions.
M.P. was supported by a grant from the Pazy Fund of the Israel Atomic Energy Commission. 

\section*{Data availability}
The data underlying this article will be shared on reasonable request to the corresponding author. 



\bibliographystyle{mnras}
\bibliography{referenses} 

\begin{thebibliography}{}
\makeatletter
\relax
\def\mn@urlcharsother{\let\do\@makeother \do\$\do\&\do\#\do\^\do\_\do\%\do\~}
\def\mn@doi{\begingroup\mn@urlcharsother \@ifnextchar [ {\mn@doi@}
  {\mn@doi@[]}}
\def\mn@doi@[#1]#2{\def\@tempa{#1}\ifx\@tempa\@empty \href
  {http://dx.doi.org/#2} {doi:#2}\else \href {http://dx.doi.org/#2} {#1}\fi
  \endgroup}
\def\mn@eprint#1#2{\mn@eprint@#1:#2::\@nil}
\def\mn@eprint@arXiv#1{\href {http://arxiv.org/abs/#1} {{\tt arXiv:#1}}}
\def\mn@eprint@dblp#1{\href {http://dblp.uni-trier.de/rec/bibtex/#1.xml}
  {dblp:#1}}
\def\mn@eprint@#1:#2:#3:#4\@nil{\def\@tempa {#1}\def\@tempb {#2}\def\@tempc
  {#3}\ifx \@tempc \@empty \let \@tempc \@tempb \let \@tempb \@tempa \fi \ifx
  \@tempb \@empty \def\@tempb {arXiv}\fi \@ifundefined
  {mn@eprint@\@tempb}{\@tempb:\@tempc}{\expandafter \expandafter \csname
  mn@eprint@\@tempb\endcsname \expandafter{\@tempc}}}

\bibitem[\protect\citeauthoryear{Dorn \& Lichtenberg}{Dorn \&
  Lichtenberg}{2021}]{Dorn2021}
Dorn C.,  Lichtenberg T.,  2021, \mn@doi [The Astrophysical Journal Letters]
  {10.3847/2041-8213/ac33af}, 922, L4

\bibitem[\protect\citeauthoryear{{Guilera}, {de El{\'\i}a}, {Brunini}  \&
  {Santamar{\'\i}a}}{{Guilera} et~al.}{2014}]{Guilera2014}
{Guilera} O.~M.,  {de El{\'\i}a} G.~C.,  {Brunini} A.,   {Santamar{\'\i}a}
  P.~J.,  2014, \mn@doi [\aap] {10.1051/0004-6361/201322061}, \href
  {https://ui.adsabs.harvard.edu/abs/2014A&A...565A..96G} {565, A96}

\bibitem[\protect\citeauthoryear{Howe, Adams  \& Meyer}{Howe
  et~al.}{2020}]{Howe2020}
Howe A.~R.,  Adams F.~C.,   Meyer M.~R.,  2020, \mn@doi [The Astrophysical
  Journal] {10.3847/1538-4357/ab620c}, 894, 130

\bibitem[\protect\citeauthoryear{Johansen, Blum, Tanaka, Ormel, Bizzarro  \&
  Rickman}{Johansen et~al.}{2014}]{Johansen2014}
Johansen A.,  Blum J.,  Tanaka H.,  Ormel C.,  Bizzarro M.,   Rickman H.,
  2014, The Multifaceted Planetesimal Formation Process.
University of Arizona Press, p. 547–570,
  \mn@doi{10.2458/azu_uapress_9780816531240-ch024}

\bibitem[\protect\citeauthoryear{{Lissauer}}{{Lissauer}}{1987}]{Lissauer1987}
{Lissauer} J.~J.,  1987, \mn@doi [Icarus] {10.1016/0019-1035(87)90104-7}, \href
  {https://ui.adsabs.harvard.edu/abs/1987Icar...69..249L} {69, 249}

\bibitem[\protect\citeauthoryear{{Lodders}}{{Lodders}}{2003}]{Lodders2003}
{Lodders} K.,  2003, \mn@doi [\apj] {10.1086/375492}, \href
  {https://ui.adsabs.harvard.edu/abs/2003ApJ...591.1220L} {591, 1220}

\bibitem[\protect\citeauthoryear{Lozovsky, Helled, Dorn  \& Venturini}{Lozovsky
  et~al.}{2018}]{Lozovsky2018}
Lozovsky M.,  Helled R.,  Dorn C.,   Venturini J.,  2018, The Astrophysical
  Journal, 866, 49

\bibitem[\protect\citeauthoryear{Lozovsky, Prialnik  \& Podolak}{Lozovsky
  et~al.}{2022}]{Lozovsky2022}
Lozovsky M.,  Prialnik D.,   Podolak M.,  2022, \mn@doi [The Astrophysical
  Journal] {10.3847/1538-4357/ac7806}, 934, 48

\bibitem[\protect\citeauthoryear{Marcus, Sasselov, Hernquist  \&
  Stewart}{Marcus et~al.}{2010}]{Marcus2010}
Marcus R.~A.,  Sasselov D.,  Hernquist L.,   Stewart S.~T.,  2010, The
  Astrophysical Journal Letters, 712, L73

\bibitem[\protect\citeauthoryear{Pollack, Hubickyj, Bodenheimer, Lissauer,
  Podolak  \& Greenzweig}{Pollack et~al.}{1996}]{Pollack1996}
Pollack J.~B.,  Hubickyj O.,  Bodenheimer P.,  Lissauer J.~J.,  Podolak M.,
  Greenzweig Y.,  1996, \mn@doi [Icarus]
  {https://doi.org/10.1006/icar.1996.0190}, 124, 62

\bibitem[\protect\citeauthoryear{{Ros} \& {Johansen}}{{Ros} \&
  {Johansen}}{2013}]{Ros2013}
{Ros} K.,  {Johansen} A.,  2013, \mn@doi [\aap] {10.1051/0004-6361/201220536},
  \href {https://ui.adsabs.harvard.edu/abs/2013A&A...552A.137R} {552, A137}

\bibitem[\protect\citeauthoryear{Safronov}{Safronov}{1972}]{safronov1972}
Safronov V.,  1972, Israel program for scientific translations

\bibitem[\protect\citeauthoryear{{Schlichting}, {Sari}  \&
  {Yalinewich}}{{Schlichting} et~al.}{2015}]{Schlichting2015}
{Schlichting} H.~E.,  {Sari} R.,   {Yalinewich} A.,  2015, \mn@doi [\icarus]
  {10.1016/j.icarus.2014.09.053}, \href
  {https://ui.adsabs.harvard.edu/abs/2015Icar..247...81S} {247, 81}

\bibitem[\protect\citeauthoryear{Vazan, Sari  \& Kessel}{Vazan
  et~al.}{2022}]{Vazan2022}
Vazan A.,  Sari R.,   Kessel R.,  2022, \mn@doi [The Astrophysical Journal]
  {10.3847/1538-4357/ac458c}, 926, 150

\makeatother
\end{thebibliography}

\bsp	
\label{lastpage}
\end{document}